\documentstyle[aps,twocolumn,prb,psfig]{revtex}
\begin{document}
\draft
\title{Anomalous ferromagnetic spin fluctuations \\ 
in an antiferromagnetic insulator Pr$_{1-x}$Ca$_{x}$MnO$_{3}$}
\author{R. Kajimoto, T. Kakeshita, Y. Oohara, H. Yoshizawa}
\address{Neutron Scattering Laboratory, I. S. S. P., University of Tokyo, Tokai, Ibaraki, 319-1106, Japan}
\author{Y. Tomioka}
\address{Joint Research Center for Atom Technology (JRCAT), Tsukuba, 305, Japan}
\author{Y. Tokura}
\address{Joint Research Center for Atom Technology (JRCAT), Tsukuba, 305, Japan  \\ 
and Department of Applied Physics, University of Tokyo, Tokyo 113-0033, Japan}

\date{\today}

\twocolumn[\hsize\textwidth\columnwidth\hsize\csname @twocolumnfalse\endcsname

\maketitle
\begin{abstract}

The high temperature paramagnetic state in an antiferromagnetic (AFM) insulator Pr$_{1-x}$Ca$_{x}$MnO$_{3}$ is characterized by the ferromagnetic (FM) spin fluctuations with an anomalously small energy scale.  The FM fluctuations show a precipitous decrease of the intensity at the charge ordering temperature $T_{\rm CO}$, but persist below $T_{\rm CO}$, and vanish at the AFM transition temperature $T_{\rm N}$.  These results demonstrate the importance of the spin ordering for the complete switching of the FM fluctuation in doped manganites.

\end{abstract}
]

For hole doped perovskite manganites R$_{1-x}$A$_{x}$MnO$_{3}$ (R is a trivalent rare earth ion, and A is a divalent alkaline earth ion.), the ferromagnetic (FM) metallic state has been qualitatively explained with the double exchange (DE) model, in which the doped holes gain the kinetic energy by aligning the spins on the Mn sites ferromagnetically.\cite{DE}  This DE picture implies that, regardless of its FM or antiferromagnetic (AFM) ordering in the ground state, the FM spin correlation must be the characteristic feature of spin fluctuations in the paramagnetic state of doped manganites.  When the charge ordering is formed, however, it will suppress the hopping of the holes, and suppresses the FM fluctuations which are assisted by the DE interactions.  In addition, the orbital ordering may turn on the AFM super exchange interactions between localized spins on the Mn sites.

Being consistent with such a DE picture, it was recently reported that the FM spin fluctuations exist in the paramagnetic phase of the insulating AFM (Bi,Ca)MnO$_{3}$, and they change over to the AFM spin fluctuations below the onset of the charge ordering. \cite{bao_bcmo}  Very recently, it is discovered that the spin correlation in the A-type AFM manganite Nd$_{0.45}$Sr$_{0.55}$MnO$_{3}$ is {\it ferromagnetic} in the paramagnetic state, and it exhibits a metallic AFM state with the orbital ordering at low temperatures.\cite{kaw97}  These recent results manifest the importance of the FM spin fluctuations in the paramagnetic state of the doped manganites.  In addition, the switching of spin fluctuations from FM to AFM correlations at the charge and/or spin ordering temperature typifies the interplay between the orbital, charge, and spin orderings in doped manganites.

In order to confirm the existence of the FM spin fluctuations in the paramagnetic state of doped manganites with an AFM ground state and to elucidate the characteristic feature of the FM spin fluctuations in the paramagnetic state, we have studied an {\it insulating AFM} system Pr$_{1-x}$Ca$_{x}$MnO$_{3}$.  We chose this material because its transport, optical, and magnetic properties are all well-characterized.\cite{jirak,tom95c,yos95,oki98}  Furthermore, this system is convenient to examine the mechanism of the switching of spin fluctuations because its well-separated charge and spin ordering temperatures,\cite{tom95c,yos95} for example, the $x= 0.35$ sample shows the CE-type charge ordering (CO) at $T_{\rm CO} \sim 230$ K, while the CE-type AFM spin ordering at $T_{\rm N} \sim 165$ K, as shown later.  Consequently, this system allows us to experimentally distinguish the influence of the spin ordering from that of the CO on the switching of spin fluctuations.

In this paper, we shall report the results of the neutron scattering studies on Pr$_{1-x}$Ca$_{x}$MnO$_{3}$ with $0.35 \leq x \leq 0.50$.  The central result of the present work is that anomalous FM fluctuations exist over a wide concentration range for $0.35 \lesssim x \lesssim 0.50$ in the insulating paramagnetic state of Pr$_{1-x}$Ca$_{x}$MnO$_{3}$ despite its AFM ground state.  The energy scale of the FM component is anomalously small when it is compared with the low temperature spin stiffness parameters.  In addition, this system shows a clear switching of the spin fluctuations from the FM spin fluctuation to the AFM one in two steps, the first change at the charge and orbital ordering temperature $T_{\rm CO}$, and the second one at the spin ordering temperature $T_{\rm N}$, demonstrating the importance of the spin ordering for the complete switching of the spin fluctuations in doped manganites.

The single crystal samples were melt grown by the floating zone method as described previously.\cite{tom95c}  The quality of the samples was checked with x-ray diffraction and with inductively coupled plasma mass spectroscopy (ICP), and they were characterized well by the transport, optical, and magnetization measurements in the previous studies.\cite{tom95c,yos95,oki98}
Neutron scattering measurements were carried out on triple-axis spectrometers HER and GPTAS in the JRR-3M of JAERI, Tokai.  An incident neutron energy of $E_{\rm i} = 4.98$ meV with a combination of open-80'-80' collimators were utilized at HER, and that of $E_{\rm i} = 13.7$ meV with a combination of 10'-40'-40'-40' collimators at GPTAS.  The energy resolution of the HER spectrometer was $E_{res} \sim 0.23$ meV (FWHM).  The samples were mounted in aluminum capsules with helium gas, and were attached to the cold head of a closed-cycle helium gas refrigerator.  The temperature of the samples was controlled within accuracy of 0.2 degrees.

\begin{figure}[htb]
\centering \leavevmode
\psfig{file=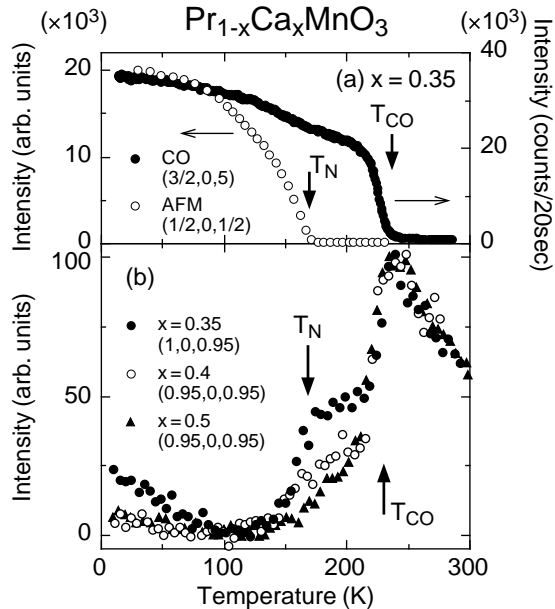,width=0.85\hsize}
\caption{Temperature dependences of order parameters for charge and AFM ordering for $x=0.35$.  (b)  Temperature dependences of FM diffuse scattering observed near (101) at $E=0$ meV.  The temperature axis is scaled at $T_{CO}$ of the $x=0.35$ sample as described in the text.}
\label{T-dep}
\end{figure}

The crystal structure of Pr$_{1-x}$Ca$_{x}$MnO$_{3}$ is orthorhombic $Pnma$ with quasi-cubic lattice constants $b/\sqrt{2} \sim a \sim c$.  For Pr$_{1-x}$Ca$_{x}$MnO$_{3}$ with $0.3 \leq x \lesssim 0.6$, the superlattice reflections of the CO are observed at $Q=(hkl)$ with $h = \mbox{half integer}$ and $k = \mbox{even}$.  On the other hand, the CE-type AFM Bragg reflections appear at $Q=(hkl)$ with $h=\mbox{half integer}$ or integer, $l=\mbox{half integer}$, and $k=\mbox{odd}$ for the CE-type $x=0.50$ sample, whereas they appear at $k=\mbox{even}$ for the {\it pseudo CE-type} magnetic structure ($x=0.35$ and 0.40) due to the FM stacking of the spins along the $b$ axis.\cite{jirak,yos95}  The FM spin correlation was observed at $Q=(hkl)$ with $h+l=\mbox{even},\,k=\mbox{even}$.  For the neutron scattering measurements, we chose the $(h0l)$ scatterng plane for the $x=0.35$ sample, while the $(hkh)$ zone for $x=0.40$ and 0.50 samples.

We first show the temperature dependence of the order parameters, and characterize the charge and spin ordering in Pr$_{1-x}$Ca$_{x}$MnO$_{3}$.  Figure \ref{T-dep}(a) is the temperature dependence of the order parameters of the charge and AFM spin order in the $x=0.35$ sample.  The CE-type charge order is established below $T_{\rm CO} = 230$ K, and the superlattice reflection is observed at $Q = (3/2, 0, 5)$.  With the further decrease of temperature, the AFM spin order is formed below $T_{\rm N} = 160$ K as is seen from the onset of the AFM Bragg intensity at $Q = (1/2, 0, 1/2)$.  The intensity at $Q = (3/2, 0, 5)$ shows a slight increase below $T_{\rm N}$ because of the superposition of the AFM component.

\begin{figure}[htb]
\centering \leavevmode
\psfig{file=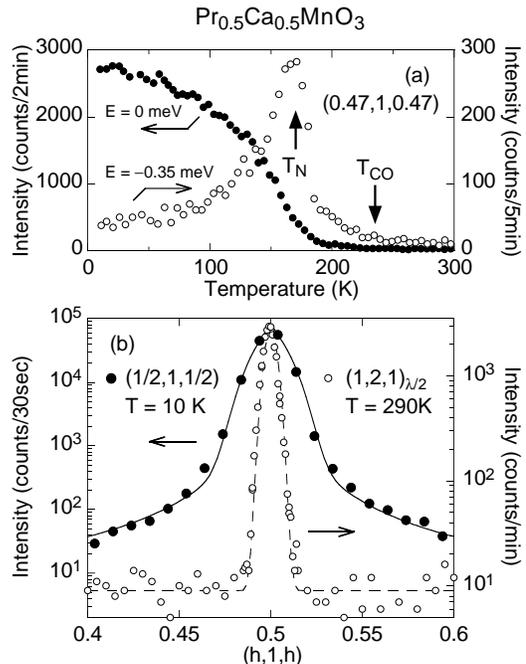,width=0.8\hsize}
\caption{(a) Temperature dependences of AFM scattering for $x=0.5$ observed near $(\frac{1}{2}1\frac{1}{2})$ at $E = 0$~meV and $E = -0.35$~meV. (b) Profile of $(\frac{1}{2} 1 \frac{1}{2})$ AFM Bragg peak at 10 K and the instrumental resolution determined by the second order reflection.}
\label{AFprof2}
\end{figure}

We have also studied behavior of the AFM spin correlations.  The data revealed that the elastic and inelastic components show quite different behavior, indicating the importance of the energy window of the observation.  The top panel of Fig. \ref{AFprof2} illustrates the temperature dependence of the AFM fluctuations in the $x=0.5$ sample measured at $E = 0$ meV and at $E = -0.35$ meV with a good energy resolution of $E_{res} \sim 0.23$ meV.  At $E = - 0.35$ meV, the dynamical AFM fluctuation exhibits an onset at $T_{\rm CO}$, and shows a sharp critical divergence around $T_{\rm N}$.  At $E = 0$ meV, on the other hand, the intensity increases monotonically similar to the AFM Bragg component shown in Fig. \ref{T-dep}(a).  This result clearly demonstrates the existence of the disordered frozen spin components with the AFM correlation in the sample, and implies the incompleteness of the CE-type AFM spin order.

To examine whether the CE-type spin order is true long range order, we measured the AFM Bragg profiles of the $x=0.5$ sample.  Surprisingly, we found that the width of the AFM Bragg peak in the $x=0.5$ sample is broader than the momentum resolution at 10 K, and exhibits a distinct wing as shown in the bottom panel of Fig. \ref{AFprof2}.  By comparing the top panel and Fig. \ref{T-dep}(a), we know the temperature dependence of the wing intensity is AFM Bragg-like.  We repeated the measurements on another 50 \% hole-doped sample, Nd$_{1/2}$Sr$_{1/2}$MnO$_{3}$, and confirmed the exactly same behavior.  These observations indicate that the CE-type spin ordering is not perfect in either of the well-characterized $x=0.5$ samples, Pr$_{1/2}$Ca$_{1/2}$MnO$_{3}$ and Nd$_{1/2}$Sr$_{1/2}$MnO$_{3}$, and lead us to conclude that the CE-type spin ordering is not true long range order even at the ideal hole concentration $n_{h} = 1/2$ for the CE-type ordering.  This incompleteness of the CE-type spin ordering causes an important effect on the temperature dependence of spin fluctuations in the CMR manganite systems as we show below.

\begin{figure}[htb]
\centering \leavevmode
\psfig{file=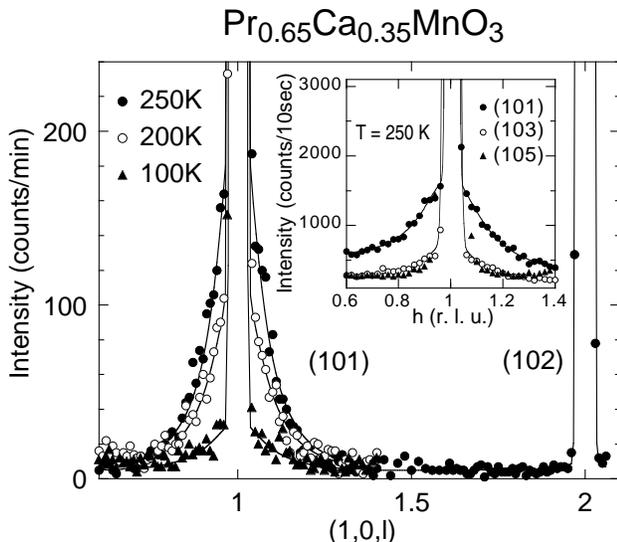,width=0.95\hsize}
\caption{Profiles of $Q=(101)$ and $(102)$ along the [001] direction at 250 K, 200 K and 100 K for $x=0.35$.  Inset: Profiles scanned along the [100] direction near $(10l)$ at 250 K.}
\label{Fprof}
\end{figure}

As we expect FM spin fluctuations in the paramagnetic state within the DE picture, the insulating AFM Pr$_{1-x}$Ca$_{x}$MnO$_{3}$ system indeed shows distinct {\it ferromagnetic} (FM) spin fluctuations in the paramagnetic state.  Figure \ref{Fprof} shows the profiles of $q$ scans along the $(1, 0, l)$ line for the $x=0.35$ sample.  In the paramagnetic phase at 250 K and 200 K, one can see strong diffuse scattering around the FM Bragg point (101), in contrast to no diffuse scattering around the nuclear Bragg point (102) at which the FM structure factor for the Mn ions vanishes.  The inset shows the scattering profiles observed at around $(1, 0, l)$ with $l$ = odd at 250 K.  The diffuse intensity at (101) is strongest, and decreases as $Q$ increases, being consistent with the magnetic form factor.  From these results, we identify that the diffuse scattering observed in the paramagnetic phase originates from the FM spin fluctuations.

Now, we return to Fig. \ref{T-dep}(b), which shows the temperature dependence of the elastic component of the FM spin fluctuations for the three samples with $x = 0.35, 0.40,$ and 0.50.  The vertical axis, the scattering intensity, is scaled so that the maximum and minimum intensity for each sample coincide.  For convenience of comparison, slightly different $T_{\rm CO}$'s of three samples ranging from 230 K to 250 K were also scaled to $T_{\rm CO} \sim$ 230 K of the $x = 0.35$ sample.  For the $x = 0.35$ sample, the FM fluctuation is suppressed clearly in two steps.  First, it suddenly decreases to half at the onset of the CO.  On the other hand, it remains finite for $T_{\rm CO} > T > T_{\rm N}$ with little temperature dependence, and then it completely vanishes at the onset of AFM spin order.  The intensity for $T_{\rm CO} > T > T_{\rm N}$ becomes weaker as the hole concentration $x$ approaches the commensurate value 1/2, but a finite amount of the intensity persists even for the $x = 0.5$ sample.

The important finding in Fig. \ref{T-dep}(b) is two successive sharp suppressions of the FM spin fluctuation at $T_{\rm CO}$ and at $T_{\rm N}$.  In principle, the FM correlation is expected to be suppressed at $T_{\rm CO}$, because the charge localization inhibits the ferromagnetic DE interactions mediated by electron hopping.  Since the FM diffuse scattering for $T_{\rm CO} \gtrsim T \gtrsim T_{\rm N}$ decreases as $x$ approaches $x =1/2$ as seen in Fig. \ref{T-dep}(b), one may consider that excess e$_{\rm g}$ electrons over the commensurate concentration mediate the FM fluctuations.  However, we have observed that the intense FM fluctuation persists below $T_{\rm CO}$ and vanishes at $T_{\rm N}$ even in the $x=0.5$ sample.  This result clearly excludes the above interpretation.  The origin of the FM diffuse component for $T_{\rm CO} \gtrsim T \gtrsim T_{\rm N}$ can be interpreted as follows.  We first note that the spin fluctuation in the paramagnetic state of the DE system is ferromagnetic when it lacks the orbital ordering.  After the system forms the orbital ordering of the e$_{\rm g}$ orbitals, the super exchange interactions actuate the AFM spin fluctuations.  Therefore, the remanence of the FM spin fluctuation below $T_{\rm CO}$ indicates the existence of the orbital fluctuations.  In the CE-type ordering, both orbitals and spins form a complicated pattern of a mixture of FM and AFM arrangements.\cite{jirak,yos95}  As a result, it may be unstable against thermal agitations, and there may exist a sufficient amount of orbital fluctuations.  The present data prove that the CE-type ordering is not true long range order, and that it allows the orbital and FM spin fluctuations for $T_{\rm CO} \gtrsim T \gtrsim T_{\rm N}$.  Consequently, the orbital pattern in the CE-type ordering is not fully stabilized even for $x = 0.5$ until the spin ordering is established at $T_{\rm N}$.

\begin{figure}[htb]
\centering \leavevmode
\psfig{file=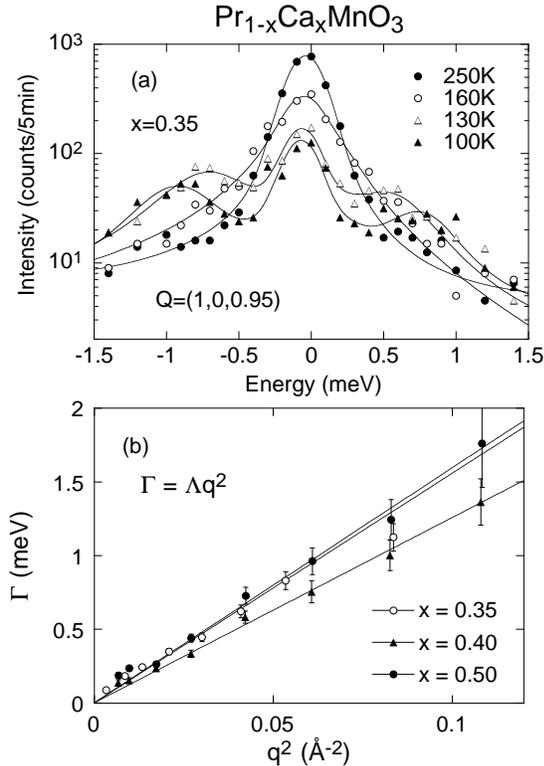,width=0.85\hsize}
\caption{(a) Energy spectra at $Q=(1,0,0.95)$ for the $x=0.35$ sample.  (b) Energy widths $\Gamma$ of quasielastic scattering versus $q^2$ above $T_{\rm CO}$; open circle: $x=0.35$ at (1,0,0.95) and 245 K, triangle: $x=0.40$ at (0.95,0,0.95) and 250 K, closed circle: $x=0.50$ at (0.95,0,0.95) and 260 K.}
\label{inela}
\end{figure}

Figure \ref{inela}(a) shows the temperature dependence of the energy spectra of the FM fluctuations in the $x=0.35$ sample.  Solid lines are the fits to the Lorentzian form convoluted with the instrumental resolution.  The intense quasielastic scattering is evident in the paramagnetic phase, and it changes over to the AFM spin wave excitations below $T_{\rm N}$.  The spin wave peaks at 160 K are too close to $E = 0$ meV, and they are not resolved from the central component.  With the decrease of temperature, however, they show a hardening, and reach to $E \sim$ 0.9 meV at 100 K.

In order to quantitatively characterize the FM spin fluctuations, we have studied the $q$ dependence just above $T_{\rm CO}$.  The quasielastic peaks were fitted to a Lorentzian, and the width $\Gamma$ is plotted against $q^2$ in Fig. \ref{inela}(b). For all three samples, $\Gamma$ shows an excellent linearity on $q^2$ ($\Gamma = \Lambda q^2$), which indicates that the scattering originates from a spin diffusion-like process. The spin diffusion constant $\Lambda$ deduced from the fits is practically $x$-independent for all three samples, yielding 16(4) meV\AA$^{2}$ (HWHM) for $x=0.35$, 13(2) meV\AA$^{2}$ for $x=0.40$, and 16(4) meV\AA$^{2}$ for $x=0.50$, repectively.  It should be noted that a similar FM component has been recently reported in the {\it metallic FM} manganites, La$_{0.67}$Ca$_{0.33}$MnO$_{3}$ and Nd$_{0.7}$Sr$_{0.3}$MnO$_{3}$,\cite{lynn,Jaime} and a {\it metallic AFM} Nd$_{0.45}$Sr$_{0.55}$MnO$_{3}$.\cite{kaw97}  We would like to point out that the energy scale $\Lambda$ of such FM fluctuations is anomalously small.  For instance, $\Lambda = 15(1)$~meV\AA$^{2}$ for La$_{0.67}$Ca$_{0.33}$MnO$_{3}$ (Ref. \onlinecite{lynn}), and $\Lambda = 14(2)$~meV\AA$^{2}$  for Nd$_{0.45}$Sr$_{0.55}$MnO$_{3}$.\cite{kaw97}

  In general, the spin wave stiffness constant $D_{SW}$ and the spin diffusion constant $\Lambda$ should be of the same order in magnitude when the same exchange interactions control the spin wave propagation as well as the spin diffusion process.  The values of $D_{SW}$ in doped manganites found in literature are of the order of 100 meV,\cite{lynn,Perr,endo,martin} and, in fact, the values $\Lambda$ of the metallic FM samples, La$_{0.8}$Sr$_{0.2}$MnO$_{3}$ and La$_{0.7}$Sr$_{0.3}$MnO$_{3}$, are reported to be of the same order with $D_{SW}$.\cite{endo,martin}  We think that the discrepancy of the energy scale between $D_{SW}$ and $\Lambda$ in Pr$_{1-x}$Ca$_{x}$MnO$_{3}$ should be attributed to its narrower one-electron bandwidth $W$.  In a system with a small $W$, the small hopping integral suppresses the mobility of holes, and increases the possibility of the coupling of the hole motion with the lattice distortion.  It should be note that the existence of the two energy scales in doped manganites is recently pointed out by studies of thermopower and resistivity measurements,\cite{jai96} and by a neutron scattering study of Nd$_{0.45}$Sr$_{0.55}$MnO$_{3}$.\cite{kaw97}  The reported activation energies of the resistivity and thermopower are quantitatively consistent with the spin wave stiffness constant $D_{SW}$ and the spin diffusion constant $\Lambda$, respectively.  It was suggested that the small energy scale of the thermopower in the paramagnetic state could be attributed to the hopping of the small polarons. \cite{jai96}

In conclusion, we have demonstrated that the paramagnetic phase in the charge ordered AFM insulator Pr$_{1-x}$Ca$_{x}$MnO$_{3}$ with $0.35 \leq x \leq 0.5$ is characterized by the anomalous FM fluctuations.  For a system with a narrow one electron bandwidth, the energy scale of the FM fluctuations is an order of magnitude smaller than that of the low temperature spin dynamics.  These results are compatible with the observation of two energy scales in the thermodynamic measurements.\cite{jai96}  The FM fluctuation is not completely suppressed at $T_{\rm CO}$, but vanishes at $T_{\rm N}$, indicating that the spin ordering is crucial to achieve the complete suppression of the orbital fluctuations and the concomitant switching of spin fluctuations.  The CE-type spin ordering is not true long range order even for the stoichiometric concentration $x=0.5$.

We thank Dr. H. Kawano for illuminating discussions and for a critical reading of the manuscript.  This work was supported by a Grant-In-Aid for Scientific Research from the Ministry of Education, Science and Culture, Japan and by the New Energy and Industrial Technology Development Organization (NEDO) of Japan.

\end{document}